\documentstyle[12pt,amssymb]{article}
\begin{document}
\tolerance=5000
\def\pp{{\, \mid \hskip -1.5mm =}}
\def\cL{{\cal L}}
\def\be{\begin{equation}}
\def\ee{\end{equation}}
\def\bea{\begin{eqnarray}}
\def\eea{\end{eqnarray}}
\def\tr{{\rm tr}\, }
\def\nn{\nonumber \\}
\def\e{{\rm e}}

\begin{titlepage}

\ \hfill IFT-P.036/2004

\begin{center}
\Large {\bf The final state and thermodynamics of dark energy universe}

\vspace*{5mm}

\normalsize

\large{Shin'ichi Nojiri\footnote{Electronic mail: nojiri@nda.ac.jp,
snojiri@yukawa.kyoto-u.ac.jp} \\
and Sergei D.Odintsov$^{\clubsuit\spadesuit}$\footnote{Electronic mail:
odintsov@ieec.fcr.es Also at TSPU, Tomsk, Russia}}

\normalsize

\vspace*{3mm}

{\em Department of Applied Physics,
National Defence Academy, \\
Hashirimizu Yokosuka 239-8686, Japan}
\medskip

{\em $\clubsuit$ Instituci\`o Catalana de Recerca i Estudis
Avan\c{c}ats (ICREA)  and
 Institut d'Estudis Espacials de Catalunya (IEEC), \\
Edifici Nexus, Gran Capit\`a 2-4, 08034 Barcelona, Spain}

{\em $\spadesuit$ Inst. Fisica Teorica, Universidade Estadual Paulista, \\
Sao Paulo, Brazil (temporary address)}
\medskip

\end{center}

\vspace*{1mm}

\begin{abstract}

As it follows from the classical analysis, the typical final state of
the dark energy universe where dominant energy condition is violated is
finite time,
sudden future singularity (Big Rip). For a number of dark energy universes
(including scalar phantom and effective phantom theories as well as
specific quintessence model) we demonstrate that quantum effects
play the dominant role near Big Rip, driving the universe out of future
singularity (or, at least, moderate it). As a consequence, the
entropy bounds with quantum corrections become well-defined near Big Rip.
Similarly, black holes mass loss due to phantom accretion is not so
dramatic as it was expected: masses do not vanish to zero due to transient
character of phantom evolution stage. Some examples of cosmological
evolution for negative, time-dependent equation of state are also
considered with the same conclusions. The application of negative entropy
(or negative temparature) occurence in the phantom thermodynamics is
briefly discussed.

\end{abstract}


\end{titlepage}

\section{Introduction}

Recent astrophysical data, ranging from WMAP observations to high
redshifts surveys of supernovae, indicate that about 70 percent of the
total energy of the universe is to be attributed to a strange cosmic fluid
with negative pressure, dark energy.
>From another side, the universe is accelerating currently. It is also
observed that the equation of state parameter $w$ is close to $-1$,
 most probably being below $-1$. (The possibility of time-dependent
negative $w$ is not excluded too).\footnote{It is quite possible that
what looks like dark energy is the manifestation of some unknown feature
of the gravitational theory which apparantly should be modified.}

The case with $w$ less than $-1$ is often dubbed as phantom dark energy.
At the moment, there is no satisfactory theoretical description of
phantom dark energy (for a number of attempts in this direction, see
\cite{phantom,phantom1} and for recent review of dark energy,
see\cite{sahni}). The easiest current model of phantom is motivated by
quintessence\cite{quint}, it is just scalar field with wrong sign
for kinetic energy term. Definitely, such theory being instable shows
some weird properties caused mainly by the violation of dominant energy
condition. Indeed, the energy density grows with time in the phantom
universe
so that in a finite time such universe ends up in the singularity dubbed
as Big Rip\cite{CKW} (see also earlier discussion of finite time
singularity in \cite{tipler}). The related phenomenon is that all black
holes loss their masses to vanish exactly in Big Rip \cite{BDE}. Phantom
thermodynamics looks also strange leading to negative entropy of the
universe \cite{BNOV} (and divergent entropies near Big Rip) or to
appearence of negative temparatures
\cite{Pedro}. If our universe is indeed phantom one, this all may call
to revision of basic physical principles governing our reality!

In the present article aiming to discuss the final state of (phantom)
dark energy universe we show that situation is much less dramatic than it
looks
from the very beginning (even in the absence of consistent phantom
theory). Indeed, with the growth of phantom energy density
the typical energies and curvature invariants grow as well. As a result,
much before Big Rip the quantum effects start to play the dominant role.
In a sense, second Quantum Gravity era begins. The simple account of the
quantum effects, in the same way as it was proposed in
refs.\cite{escape,emilio},
demonstrates that Big Rip singularity is moderated or even does not
occur at all. As a result, the entropy bounds remain to be meaningful and
black
holes masses do not vanish to zero. This observation indicates also
that phantom stage (if it is realistic) is just transient period in the
universe evolution.

The paper is organized as follows. In the next section we start from
the finite time, sudden singularity model proposed by Barrow \cite{barrow}
 and consider its generalization and give its lagrangian description in terms
of scalar-tensor theory. In fact, in terms of scalar-tensor theory 
the sudden singularity is different from the model of \cite{barrow}.
It is interesting that such model where dominant energy condition is
also violated is not necessary phantom with wrong sign for kinetic term.
Then we show that the account of quantum effects (using quantum energy
density and pressure obtained by integration of the conformal anomaly)
  moderates the finite time singularity  or even prevents it.
The universe presumably ends up in deSitter phase (future inflationary
era). 
A similar analysis has been done in \cite{escape} but with coefficient 
of one of the terms in conformal anomaly ($\Box R$) being equal to zero.
In the present paper, due to the importance of the corresponding term 
at high energies (as being advocated by Hawking) the account of arbitrary 
such coefficient is made. Moreover, 
 the backreaction from the matter 
 is included.  
Section three is devoted to the study of final stage for scalar
phantom universe and effective phantom universe \cite{LR} produced by
higher
derivative coupling of scalar kinetic energy with curvature. Again,
quantum escape of Big Rip occurs or, at least, the singularity
 is moderating permitting the evolution after Big Rip time.
In section four the entropy bounds near Big Rip are studied.
Basically, the typical entropies (including the one for negative
time-dependent equation of state universe) diverge at singularity.
The account of quantum effects makes the entropies to be finite and
the entropy bounds to be well-behaved.
Section five is devoted to study of more general, time-dependent equation
of state which may be also effectively phantom. The examples where scale
factor is accelerating are presented and the occurence of Big Rip is
mentioned again. Finally, some summary and outlook are given
in Discussion. In Appendix A the entropy is written for specific
model of phantom thermodynamics. It becomes negative for positive
temperatures, and positive if temperatures are negative.
In  Appendix B the evolution of black holes mass
in phantom universe is discussed, also in the case when equation of state
is time-dependent. The same quantum effects which drive the final state
out of Big Rip significally improve the evolution of black holes mass.
It may increase or decrease by phantom energy accretion but
eventually does not vanish. 

\section{Lagrangian description of classical sudden future singularity and
quantum effects account}

In this section, we construct the  scalar-tensor theory with specific
potential which describes classical sudden future singularity. The account
of
quantum effects near to sudden singularity (where future quantum gravity
era starts) shows that sudden singularity most probably never occurs.

In \cite{barrow}, it has been shown that even if the strong energy condition
\be
\label{B0}
\rho>0 \ ,\quad \rho + 3p>0
\ee
for some kind of (exotic) matter is satisfied,
the future singularity ( Big Rip) can occur. Note that dominant energy
condition is violated \cite{lake} for such a scenario.
Here $\rho$ and $p$ are the energy density and the pressure of the matter, respectively.
We now consider the spatially-flat FRW metric
\be
\label{dSP2}
ds^2= -dt^2 + a(t)^2\sum_{i=1,2,3} \left(dx^i\right)^2\ .
\ee
Following \cite{barrow},  the scale factor $a(t)$
is chosen as
\be
\label{B1}
a(t)=A + B t^q + C\left(t_s - t\right)^n\ .
\ee
Here $A>0$, $B>0$, $q>0$, and $t_s>0$ are constants and $C=-At_s^{-n}$.
 It is assumed that
$t<t_s$ and $2>n>1$. There is a singularity at $t\to t_s$, where $\frac{1}{a}
\frac{d^2 a}{dt^2}\to + \infty$.
Classical FRW equations
\be
\label{B2}
\frac{6}{\kappa^2}H^2 =\rho\ ,\quad \frac{2}{\kappa^2 a}\frac{d^2 a}{dt^2}
= - \frac{\rho + 3p}{6}\ ,
\ee
show that
\be
\label{B3}
\rho \sim \frac{6q^2 B^2 t_s^{2q-2} }{\kappa^2 \left(A + B t_s^q\right)}>0\ ,\quad
p \sim -\frac{Cn(n-1)\left(t_s - t\right)^{n-2}}{A+B t_s^q}>0\ ,
\ee
near the singularity $t\sim t_s$.
In (\ref{B2}),  $H\equiv \frac{1}{a}\frac{da}{dt}$.
Thus, the energy density $\rho$ is finite but the pressure $p$ diverges.
Nevertheless, the strong energy condition (\ref{B0}) is satisfied since
$\rho$ and $p$ are positive.

Let us present the generalization of Barrow's model \cite{barrow} as it
was presented in ref.\cite{escape}.
In this model,
the matter has been given implicitly via the FRW equations:
\be
\label{F1}
\rho=\frac{6}{\kappa^2}H^2\ ,\quad p=-\frac{2}{\kappa^2}\left(2\frac{dH}{dt} + 3H^2\right)\ .
\ee
One may assume $H$ has the following form:
\be
\label{F2}
H(t)=\tilde H(t) + {A'}\left|t_s - t\right|^\alpha\ .
\ee
Here $\tilde H$ is a smooth, differentiable (infinite number of times, in principle) function and ${A'}$ and $t_s$
are constants.
Another assumption is that a constant $\alpha$ is not a positive integer. Then $H(t)$ has a singularity at $t=t_s$.
In case $\alpha$ is negative integer, the singularity is pole. Even if $\alpha$ is positive, in case $\alpha$ is not
an integer, there appears  a singularity, 
that is, if we analytically continue $t$ to the region $t>t_s$ from the region $t<t_s$, $H$ might 
become complex and double-valued due to the cut which appears when we analytically continue $t$ to be 
a complex number.
It is important for us that  singularity presents.
If we can consider the region $t>t_s$, there is no finite-time {\it future} singularity.

When $\alpha>1$, one gets
\be
\label{F3}
\rho \sim -p \sim \frac{6}{\kappa^2}{\tilde H\left(t_s\right)}^2\ .
\ee
Hence $w=\frac{p}{\rho}=-1$, which may correspond to the positive cosmological constant.

The case $0<\alpha<1$ corresponds to Barrow's model and when $t\sim t_s$, we find
\be
\label{F4}
\rho \sim \frac{6}{\kappa^2}{\tilde H\left(t_s\right)}^2\ ,\quad
p\sim \pm \frac{4{A'}\alpha}{\kappa^2}\left|t_s - t\right|^{\alpha-1}\ .
\ee
Here the plus sign in $\pm$ corresponds to $t<t_s$ case and the minus one to $t>t_s$.
In the following, the upper (lower) sign always corresponds to $t<t_s$ ($t>t_s$).
The  parameter of equation of state $w$ is given by
\be
\label{F5}
w=\pm \frac{2}{3}\frac{{A'}\alpha \left|t_s - t\right|^{\alpha-1}}{\tilde H\left(t_s\right)^2} \ .
\ee
Hence, $w$ is positive in two cases: one is ${A'}>0$ and $t<t_s$, which
directly corresponds
to Barrow's model,  and another is ${A'}<0$ $t>t_s$. In other cases, $w$ is negative.

When $-1<\alpha<0$, the energy density $\rho$ and the pressure is given by
\be
\label{F6}
\rho=\frac{6{A'}^2}{\kappa^2}\left|t_s - t\right|^{2\alpha}\ ,\quad
p\sim \pm \frac{4{A'}\alpha}{\kappa^2}\left|t_s - t\right|^{\alpha-1}\ .
\ee
The parameter of equation of state is
\be
\label{F7}
w=\pm \frac{2\alpha}{3{A'}}\left|t_s - t\right|^{-\alpha-1}\ ,
\ee
which diverges at $t=t_s$. Here $w$ is positive when ${A'}>0$ and $t<t_s$ or ${A'}<0$ and $t>t_s$.
The former case corresponds to sudden future singularity even if $w$ is
positive.
The singularity can be regarded as
a Big Rip. (For the recent comparison of phantom Big Rip with above type
of it, see \cite{ruth}.) The account of quantum effects leads to the
escape from
the future singularity as it was shown in refs.\cite{escape,emilio}.

The case $\alpha=-1$ gives
\be
\label{F8}
\rho=\frac{6{A'}^2}{\kappa^2}\left|t_s - t\right|^{-2}\ ,\quad
p\sim -\frac{2}{\kappa^2}\left(\pm 2{A'} + 3{A'}^2\right)\left|t_s - t\right|^{-2}\ ,
\ee
which may correspond to the scalar field with exponential potential. The parameter $w$ is given by
\be
\label{F9}
w=-1\mp \frac{2}{{A'}}\ .
\ee
Near $t=t_s$, the universe is expanding if ${A'}>0$ and $t<t_s$ or ${A'}<0$ and $t>t_s$.
The former case corresponds to the phantom with $w<-1$. In the latter case,
 if $2>{A'}>0$, the equation of state describes the usual matter with positive $w$ and if ${A'}>2$,
the matter may be the quintessence with $0>w>-1$.

If  $\alpha<-1$, one obtains
\be
\label{F9b}
\rho=-p=\frac{6{A'}^2}{\kappa^2}\left|t_s - t\right|^{2\alpha}\ ,
\ee
which gives $w=-1$ as for the cosmological constant case. In this case, however, there is sharp
 singularity at $t=t_s$ 
since both of $\rho$ and $p$ diverge at $t=t_s$. 
 This is contrary to the case $\alpha>-1$  as 
they  tend to infinity in the limit $t\to t_s$ more rapidly than in case $\alpha\geq -1$.


Let us construct the Lagrangian (scalar-tensor) model which contains
sudden future singularity.
We start from  rather general action of scalar field $\phi$ coupled with gravity:
\be
\label{R1}
S={1 \over \kappa^2}\int d^4 x \sqrt{-g} \left( R - {1 \over 2}
\partial_\mu \phi \partial^\mu \phi - V(\phi) \right)\ .
\ee
Then the energy density $\rho$ and the pressure $p$ are given by
\be
\label{BaL1}
\rho=\frac{1}{2}\left(\frac{d\phi}{dt}\right)^2 + V(\phi)\ ,\quad
p=\frac{1}{2}\left(\frac{d\phi}{dt}\right)^2 - V(\phi)\ .
\ee
The scalar equation of motion is
\be
\label{BaL9}
0=\frac{d^2\phi}{dt^2} + 3H\frac{d\phi}{dt} + V'(\phi)\ .
\ee
We are searching for the potential $V(\phi)$, which gives a solution
\be
\label{BaL13}
H=h_0 + h_1 \left(t_s -t\right)^\alpha\ ,\quad \phi=\phi_0 \left(t_s - t\right)^\beta
\ee
with constants $h_0$, $h_1$, $\alpha$, $\phi_0$, and $\beta$.
Since
\bea
\label{BaL14}
\frac{d\phi}{dt}&=& - \beta\phi_0\left(t_s - t\right)^{\beta -1}=- \beta\phi_0^\frac{1}{\beta}
\phi^{1-\frac{1}{\beta}}\ ,\nn
\frac{d^2\phi}{dt^2}&=& \beta\left(\beta - 1\right)\phi_0\left(t_s - t\right)^{\beta -2}
=\beta\left(\beta -1 \right) \phi_0^\frac{2}{\beta}
\phi^{1-\frac{2}{\beta}}\ ,\nn
H&=&h_0 + h_1\phi_0^{-\frac\alpha\beta}\phi^\frac\alpha\beta\ .
\eea
>From the FRW equation
\be
\label{BaL15}
\frac{6}{\kappa^2}H^2 = \frac{1}{2}\left(\frac{d\phi}{dt}\right)^2 + V\left(\phi\right)\ ,
\ee
it follows
\bea
\label{BaL16}
V\left(\phi\right)&=& \frac{6}{\kappa^2}H^2 - \frac{1}{2}\left(\frac{d\phi}{dt}\right)^2 \nn
&=&\frac{6}{\kappa^2}\left(h_0^2 + 2h_0 h_1 \phi_0^{-\frac\alpha\beta}\phi^\frac\alpha\beta
+ h_1^2\phi_0^{-\frac{2\alpha}\beta}\phi^\frac{2\alpha}\beta\right)
 - \frac{\beta^2}{2}\phi_0^\frac{2}{\beta} \phi^{2-\frac{2}{\beta}}\ ,
\eea
which gives
\be
\label{BaL17}
V'\left(\phi\right)=\frac{12\alpha}{\kappa^2\beta }\left(
h_0 h_1 \phi_0^{-\frac\alpha\beta}\phi^{\frac\alpha\beta -1}
+ h_1^2\phi_0^{-\frac{2\alpha}\beta}\phi^{\frac{2\alpha}\beta-1}\right)
 - \left(\beta-1\right)\beta\phi_0^\frac{2}{\beta} \phi^{1-\frac{2}{\beta}}\ .
\ee
On the other hand, from the scalar equation of
motion:
\be
\label{BaL18}
0=\frac{d^2\phi}{dt^2} + 3H\frac{d\phi}{dt} + V'\left(\phi\right)\ ,
\ee
one obtains
\bea
\label{BaL19}
V'\left(\phi\right) &=& - \frac{d^2\phi}{dt^2} - 3H\frac{d\phi}{dt} \nn
&=& - \left(\beta-1\right)\beta\phi_0^\frac{2}{\beta} \phi^{1-\frac{2}{\beta}}
+ 3h_0\beta \phi_0^\frac{1}\beta\phi^{1 - \frac{1}\beta }
+ 3h_1\beta\phi_0^{-\frac\alpha\beta + \frac{1}\beta}\phi^{1+\frac\alpha\beta - \frac{1}\beta}\ .
\eea
Comparing (\ref{BaL17}) and (\ref{BaL19}), we get
\be
\label{BaL20}
\beta=\frac{\alpha + 1}2\ ,\quad h_1=\frac{\kappa^2\left(\alpha + 1 \right)^2 \phi_0^2}{16\alpha}\ .
\ee
$h_0$ can be arbitrary. Then by substituting (\ref{BaL20}) into (\ref{BaL16}), we obtain
\bea
\label{BaL21}
V(\phi)&=&\frac{6h_0^2}{\kappa^2} + \frac{3h_0\left(\alpha + 1\right)^2}{4\alpha}
\phi_0^\frac{2}{\alpha + 1}\phi^\frac{2\alpha}{\alpha + 1} + \frac{3\kappa^2\left(\alpha + 1\right)^4}{128\alpha^2}
\phi_0^\frac{4}{\alpha + 1}\phi^\frac{4\alpha}{\alpha + 1} \nn
&& - \frac{\left(\alpha + 1\right)^2}8 \phi_0^\frac{4}{\alpha + 1}\phi^{2\left(\frac{\alpha - 1}{\alpha + 1}\right)}\ .
\eea

It is interesting to investigate the (in)stability of the scalar theory
(\ref{R1}) with potential (\ref{BaL21}).
The perturbations from the solution  (\ref{BaL13}) are:
\be
\label{BaL22}
H=h_0 + h_1 \left(t_s -t\right)^\alpha + \delta h \ ,\quad \phi=\phi_0 \left(t_s - t\right)^{\frac{\alpha+1}{2}} + \delta \phi\ .
\ee
Then from the FRW equation (\ref{BaL15}) and $\phi$-equation
(\ref{BaL18}), it follows
\bea
\label{BaL23}
0&=&-\frac{12}{\kappa^2}H_0\delta h + \frac{d\Phi_0}{dt}\frac{d\delta\phi}{dt} + V'\left(\Phi_0\right)\delta\phi\ ,\\
\label{BaL24}
0&=& \frac{d^2\delta\phi}{dt^2} + 3H_0\frac{d\delta\phi}{dt} + 3\delta h \frac{d\Phi_0}{dt}
+ V''\left(\Phi_0\right)\delta \phi\ .
\eea
Here
\be
\label{BaL25}
H_0\equiv h_0 + h_1 \left(t_s -t\right)^\alpha \ ,\quad \Phi_0=\phi_0 \left(t_s - t\right)^{\frac{\alpha+1}{2}} \ .
\ee
Since
\bea
\label{BaL26}
V'\left(\Phi_0\right)&=& \frac{3(\alpha+1)h_0\phi_0}{2}\left(t_s - t\right)^{\frac{\alpha-1}{2}}
+ \frac{3\kappa^2(\alpha + 1)^3\phi_0^3}{32\alpha}\left(t_s - t\right)^{\frac{3\alpha-1}{2}} \nn
&& - \frac{\left(\alpha^2 - 1\right)\phi_0}{4}\left(t_s - t\right)^{\frac{\alpha-3}{2}} \ ,\nn
V''\left(\Phi_0\right)&=& \frac{3(\alpha-1)h_0}{2}\left(t_s - t\right)^{-1}
+ \frac{3\kappa^2(\alpha + 1)^2(3\alpha - 1)\phi_0^2}{32\alpha}\left(t_s - t\right)^{\alpha-1} \nn
&& - \frac{\left(\alpha - 1\right)(\alpha - 3)}{4}\left(t_s - t\right)^{-2} \ .
\eea
Hence, when $t\to t_s$, if $\alpha>-1$, the third terms of
$V'\left(\Phi_0\right)$ and $V''\left(\Phi_0\right)$
dominate and if $\alpha<-1$, the second terms dominate.
For the case of  Barrow model,  $0<\alpha<1$. When $t\sim
t_s$,
Eqs.(\ref{BaL23}) and (\ref{BaL24}) are:
\bea
\label{BaL27}
0&\sim & - \frac{12}{\kappa^2}h_0\delta h - \frac{(\alpha + 1)\phi_0}{2}\left(t_s - t\right)^{\frac{\alpha-1}{2}}
\frac{d\delta\phi}{dt} \nn
&& - \frac{\left(\alpha^2 - 1\right)\phi_0}{4}\left(t_s - t\right)^{\frac{\alpha-3}{2}}
\delta\phi \\
\label{BaL28}
0&\sim& \frac{d^2\delta\phi}{dt^2} + 3h_0\frac{d\delta\phi}{dt} - \frac{3(\alpha+1)\phi_0}{2}
\left(t_s - t\right)^{\frac{\alpha-1}{2}}\delta h  \nn
&& - \frac{(\alpha - 1)(\alpha - 3)}{4}\left(t_s - t\right)^{-2}\delta \phi\ .
\eea
By deleting $\delta h$ from (\ref{BaL27}) and (\ref{BaL28}), we obtain
\be
\label{BaL29}
0\sim \frac{d^2\delta\phi}{dt^2}  - \frac{(\alpha - 1)(\alpha - 3)}{4}
\left(t_s - t\right)^{-2}\delta \phi\,.
\ee
Its solution is given by
\be
\label{BaL30}
\delta\phi = \phi_1\left(t_s - t\right)^{\frac{\alpha-1}{2}}
+ \phi_2\left(t_s - t\right)^{\frac{3 -\alpha}{2}} \ .
\ee
Here $\phi_1$ and $\phi_2$ are constants.
Hence,
\be
\label{BaL31}
\delta h = - \frac{6(\alpha + 1)(\alpha -2)}{\kappa^2 h_0}\phi_2\ .
\ee
In this order of the perturbation, $\phi_1$ does not appear in $\delta h$.
Since, of course, $\frac{\alpha - 1}{2}<\frac{\alpha + 1}{2}$, when $t\to t_s$,
the first term in $\delta\phi$ (\ref{BaL30}) becomes large more rapidly than the unperturbative part
$\Phi_0=\phi_0 \left(t_s - t\right)^{\frac{\alpha+1}{2}}$
(\ref{BaL24}), which tells that the solution (\ref{BaL13}) which describes the 
 sudden singularity model in scalar-tensor theory is not stable. Already on the classical
level, such instability may stop the appearence of future singularity.
However, more secure mechanism which  acts against the singularity
occurence is quantum effects account.

Near the singularity at $t=t_s$, the curvature becomes large in general.
As the quantum corrections
usually contain the powers of the curvature ( higher derivative terms),
the correction becomes
important near the singularity. One may include the quantum effects by
taking
into account the conformal anomaly contribution as back-reaction near the
singularity.
The conformal anomaly $T_A$ has the following form:
\be
\label{OVII}
T_A=b\left(F+{2 \over 3}\Box R\right) + b' G + b''\Box R\ ,
\ee
where $F$ is the square of 4d Weyl tensor, $G$ is Gauss-Bonnet invariant.
In general, with $N$ scalar, $N_{1/2}$ spinor, $N_1$ vector fields, $N_2$ ($=0$ or $1$)
gravitons and $N_{\rm HD}$ higher derivative conformal scalars, $b$, $b'$ and $b''$ are
given by
\bea
\label{bs}
&& b={N +6N_{1/2}+12N_1 + 611 N_2 - 8N_{\rm HD} \over 120(4\pi)^2}\nn
&& b'=-{N+11N_{1/2}+62N_1 + 1411 N_2 -28 N_{\rm HD} \over 360(4\pi)^2}\ .
\eea
As is seen $b>0$ and $b'<0$ for the usual matter except the higher
derivative conformal scalars.
Notice that $b''$ can be shifted by the finite renormalization of the
local counterterm $R^2$, so  $b''$ can be arbitrary (in ref.\cite{escape}
it was chosen to be zero, for simplicity).
In terms of the corresponding  energy density $\rho_A$ and pressure $p_A$,
$T_A$ is given by $T_A=-\rho_A + 3p_A$.
Then by using the energy conservation law in  FRW universe
\be
\label{CB1}
0=\frac{d\rho_A}{dt} + 3 H\left(\rho_A + p_A\right)\ ,
\ee
we may delete $p_A$ as
\be
\label{CB2}
T_A=-4\rho_A - \frac{1}{H}\frac{d\rho_A}{dt}\ ,
\ee
which gives the following expression for $\rho_A$:
\bea
\label{CB3}
\rho_A&=& -\frac{1}{a^4} \int dt a^4 H T_A \nn
&=&  -\frac{1}{a^4} \int dt a^4 H \left[ -12b \left(\frac{dH}{dt}\right)^2 + 24b'\left\{
 - \left(\frac{dH}{dt}\right)^2 + H^2 \frac{dH}{dt} + H^4\right\} \right. \nn
 && \left. - 6\left(\frac{2}{3}b + b''\right)\left\{ \frac{d^3H}{dt^3} + 7 H \frac{d^2 H}{dt^2}
+ 4 \left(\frac{dH}{dt}\right)^2 + 12 H^2 \frac{dH}{dt}\right\}\right]\ .
\eea
Moreover,
\bea
\label{CB4}
p_A&=& - \rho_A - \frac{1}{3H}\frac{d\rho_A}{dt} \nn
&=& \frac{T_A}{3}  -\frac{1}{a^4} \int dt a^4 H T_A \nn
&=&  \frac{1}{3}\left[ -12b \left(\frac{dH}{dt}\right)^2 + 24b'\left\{
 - \left(\frac{dH}{dt}\right)^2 + H^2 \frac{dH}{dt} + H^4\right\} \right. \nn
 && \left. -  6\left(\frac{2}{3}b + b''\right)\left\{ \frac{d^3H}{dt^3} + 7 H \frac{d^2 H}{dt^2}
+ 4 \left(\frac{dH}{dt}\right)^2 + 12 H^2 \frac{dH}{dt}\right\}\right] \nn
&& -\frac{1}{a^4} \int dt a^4 H \left[ -12b \left(\frac{dH}{dt}\right)^2 + 24b'\left\{
 - \left(\frac{dH}{dt}\right)^2 + H^2 \frac{dH}{dt} + H^4\right\} \right. \nn
 && \left. - 6\left(\frac{2}{3}b + b''\right)\left\{ \frac{d^3H}{dt^3} + 7 H \frac{d^2 H}{dt^2}
+ 4 \left(\frac{dH}{dt}\right)^2 + 12 H^2 \frac{dH}{dt}\right\}\right]\ .
\eea

As in (\ref{BaL13}), one assumes
\be
\label{BaL32}
H\sim h_0' + h_1' \left(t_s -t\right)^{\alpha'}\ ,\quad \mbox{or} \quad a=a_0\e^{h_0't  -  \frac{h_1'}{\alpha' + 1}
\left(t_s - t\right)^{\alpha' - 3}}\ .
\ee
We also consider the case $t\sim t_s$ and  keep only the first and
 the last terms in $V(\phi)$
(\ref{BaL21})
\be
\label{BaL33}
V(\phi)\sim \frac{6h_0^2}{\kappa^2} - \frac{\left(\alpha + 1\right)^2}8
\phi_0^\frac{4}{\alpha + 1}\phi^{2\left(\frac{\alpha - 1}{\alpha + 1}\right)}\ .
\ee
These terms are dominant for Barrow model $0<\alpha<1$. Then the
consistent solution is given when $\alpha'>2$. From
(\ref{BaL9}), it is seen the behavior of
$\phi$ is not so changed from the classical solution  (\ref{BaL13})
\be
\label{BaL34}
\phi\sim \phi_0 \left(t_s - t\right)^\beta\ .
\ee
Now the quantum corrected FRW equations are:
\bea
\label{BaL35}
0&=& -\frac{6}{\kappa^2}H^2 + \frac{1}{2}\left(\frac{d\phi}{dt}\right)^2 + V(\phi) + \rho_A\ ,\\
\label{BaL36}
0&=&\frac{2}{\kappa^2}\left(2\frac{dH}{dt} + 3H^2\right) +
\frac{1}{2}\left(\frac{d\phi}{dt}\right)^2 - V(\phi) + p_A\ .
\eea
Substituting (\ref{BaL32}) into (\ref{CB3}), we obtain
\be
\label{CB5}
\rho_A \sim -24b'{h_0'}^5 \e^{-4h_0't}\int dt \e^{4h_0't}
= - 6b'{h_0'}^4 + \rho_{A0}\ .
\ee
Here $\rho_{A0}$ is  the integration constant which may be chosen to be
zero   since $\rho_A\to 0$ when
$b'\to 0$ (classical limit).
Substituting (\ref{BaL34}) and (\ref{CB5}) with $\rho_{A0}=0$ into
(\ref{BaL35}), we obtain
\be
\label{BaL37}
0=-b'\kappa^2 {h_0'}^4 - {h_0'}^2 + h_0^2\ ,
\ee
which can be solved as
\be
\label{BaL38}
{h_0'}^2 = \frac {1\pm\sqrt{1+4b'\kappa^2 h_0^2}}{-2b'\kappa^2}>0\ ,
\ee
if
\be
\label{BaL39}
1 + 4b'\kappa^2 h_0^2 \geq 0\ ,
\ee
which gives a non-trivial constraint since $b'<0$ in general.
In (\ref{BaL38}), the minus sign in $\pm$ corresponds to the classical case (\ref{BaL13}) in the
limit of $b'\to 0$. In (\ref{BaL36}), when $t\sim t_s$, one finds
\be
\label{BaL40}
p= \frac{1}{2}\left(\frac{d\phi}{dt}\right)^2 - V(\phi) \sim \frac{(\alpha+1)^2 \phi_0^2}{4}
\left(t_s - t\right)^{\alpha-1}\ .
\ee

First interesting case is that ${2 \over 3}b + b''$ does not vanish.
Since
\bea
\label{BaL41}
p_A&\sim & -2 \left({2 \over 3}b + b''\right)\frac{d^3 H}{dt^3} \nn
&\sim & 2 \left({2 \over 3}b + b''\right)\alpha'\left(\alpha' -1\right)\left(\alpha' - 2\right)h_1'
\left(t_s - t\right)^{\alpha'-3}\ ,
\eea
the classical term $\frac{2}{\kappa^2}\left(2\frac{dH}{dt}+ 3H^2\right) $ in (\ref{BaL36}) can
be neglected as this term behaves as $\left(t_s - t\right)^{\alpha'-1}$.
Due to $p\sim -p_A$, one gets
\be
\label{BaL41b}
\alpha'=\alpha + 2\ ,\quad h_1' = - \frac{(\alpha + 1)\phi_0^2}{8\left({2 \over 3}b + b''\right)(\alpha + 2)\alpha}\ .
\ee
As
\be
\label{BaL42}
H\sim h_0' + h_1' \left(t_s -t\right)^{\alpha + 2}\ ,
\ee
the singularity at $t=t_s$ is moderated,
that is, the exponent of the power of $t_s -t$ becomes larger.
When ${2 \over 3}b + b''>0$, near $t=t_s$, $H$ decreases with time, that is,
 the universe is deccelerating.
On the other hand, when  ${2 \over 3}b + b''<0$, $H$ increases with time,
 that is, the universe is
accelerating. If we may replace $\left(t_s-t\right)$ with its absolute value
 $\left|t_s - t\right|$, the deccelerating
(accelerating) universe turns to accelerate (deccelerate) when $t>t_s$.


Another interesting situation corresponds to ${2 \over 3}b + b''=0$ by
properly
choosing $b''$. In this case
\be
\label{BaL43}
p_A \sim -4\left(b-2b'\right)\left(\frac{dH}{dt}\right)^2 \sim
-4\left(b-2b'\right){h_1'}^2 {\alpha'}^2\left(t_s - t\right)^{2\left(\alpha' - 1\right)}\ .
\ee
The choice consistent with (\ref{BaL36}) is
\be
\label{BaL44}
\alpha'=\frac{\alpha+1}{2} \ ,\quad
{h_1'}^2 = - \frac{\phi_0^2}{ 4\left( b - 2b'\right)}\ .
\ee
Since $\alpha' - \alpha = \frac{1-\alpha}{2}>0$, the singularity is moderated, compared with
the classical case (\ref{BaL13}).

One may also consider the case that the classical energy density $\rho$
and the pressure $p$ can be
neglected since the quantum induced  $\rho_A$ and $p_A$ become
significally dominant. In this case,
combining the first FRW equation and (\ref{CB3})
\bea
\label{CB6}
&& \frac{6}{\kappa^2}H^2 =  -\frac{1}{a^4} \int dt a^4 H \left[ -12b \left(\frac{dH}{dt}\right)^2 + 24b'\left\{
 - \left(\frac{dH}{dt}\right)^2 + H^2 \frac{dH}{dt} + H^4\right\} \right. \nn
 && \qquad \left. - 6\left(\frac{2}{3}b + b''\right)\left\{ \frac{d^3H}{dt^3} + 7 H \frac{d^2 H}{dt^2}
+ 4 \left(\frac{dH}{dt}\right)^2 + 12 H^2 \frac{dH}{dt}\right\}\right]\ ,
\eea
one has
\bea
\label{CB7}
\lefteqn{\frac{12}{\kappa^2}\left(2H^2 + \frac{dH}{dt}\right)} \nn
&=& - \left[ -12b \left(\frac{dH}{dt}\right)^2 + 24b'\left\{
 - \left(\frac{dH}{dt}\right)^2 + H^2 \frac{dH}{dt} + H^4\right\} \right. \nn
 && \left. - 6\left(\frac{2}{3}b + b''\right)\left\{ \frac{d^3H}{dt^3} + 7 H \frac{d^2 H}{dt^2}
+ 4 \left(\frac{dH}{dt}\right)^2 + 12 H^2 \frac{dH}{dt}\right\}\right]\ ,
\eea
Notice Eq.(\ref{CB7}) is nothing but
\be
\label{CB8}
\frac{2}{\kappa^2}R=-T_A\ .
\ee
Eq.(\ref{CB7}) has a special solution, which gives a deSitter space
 with constant $H$. In fact,
if  $H$ is assumed to be the constant, Eq.(\ref{CB9}) reduces to
\be
\label{CB9}
\frac{24}{\kappa^2}H^2=-24 b' H^4\ ,
\ee
which has solutions
\be
\label{CB10}
H^2=0\ ,\quad H^2 = -\frac{1}{b'\kappa^2}\ .
\ee
The second solution describes  deSitter space.

If the curvature becomes significally large, one may neglect the classical
part, which is the l.h.s.
of (\ref{CB7}). Assuming
\be
\label{CB11}
H\sim \frac{h_0}{t}\ ,
\ee
one arrives at the following algebraic equation:
\bea
\label{CB12}
0&=&12 h_0\left\{-3\left(\frac{2}{3}b + b''\right) + \left( - b - 2b' + 9\left(\frac{2}{3}b + b''\right) \right)h_0
\right. \nn
&& \left. + \left( -2b' - 6\left(\frac{2}{3}b + b''\right) \right)h_0^2 + 2b' h_0^3\right\}\ .
\eea
Since the part inside $\{\ \}$ of (\ref{CB12}) is the third order polynomial, there is always a nontrivial
solution for $h_0$, at least, if  $\frac{2}{3}b + b''$ does not vanish.
If the obtained $h_0$ is negative, the universe is shrinking but if we change the direction
of the time by $T\to t_s - t$, we may obtain a solution describing the expanding universe.
Even if $\frac{2}{3}b + b''=0$, one gets non-trivial (non-vanishing)
solution for $h_0$:
\be
\label{CB13}
h_0=\frac{1}{2}\pm \sqrt{\frac{5}{4} + \frac{b}{2b'}}\ .
\ee
Since $b''$ is arbitrary in principle, we may consider the case that the terms with $b''$ become dominant.
Then Eq.(\ref{CB7}) reduces to
\be
\label{CB14}
\frac{12}{\kappa^2}\left(2H^2 + \frac{dH}{dt}\right)
= - 6 b''\left\{ \frac{d^3H}{dt^3} + 7 H \frac{d^2 H}{dt^2}
+ 4 \left(\frac{dH}{dt}\right)^2 + 12 H^2 \frac{dH}{dt}\right\}\ ,
\ee
which can be written by using the scalar curvature $R=6\left(2H^2 + \frac{dH}{dt}\right)$ as
\be
\label{CB15}
\frac{2}{\kappa^2}R=-b''\left(\frac{d^2 R}{dt^2} + 3H \frac{dR}{dt}\right)\ .
\ee
Eq.(\ref{CB15}) has been found in $R^2$-gravity \cite{Str,MMS} with the
purpose to describe the inflation. Thus, like in ref.\cite{escape}
(where $b''$ was chosen to be zero) we come to the following picture.
Near to future singularity, the quantum effects become dominant and they
drive (most probably) the universe to deSitter space. Thus, final state
of such universe is not the singularity. Rather, far in future the new
inflation era (which is supported by quantum gravity effects\cite{emilio})
starts.

\section{Final state of dark energy universe}

In the same way the singularity avoidance in other models (of dark energy
universe) may be considered.
First of all, let us give the simple argument stressing that Big Rip
should not occur. Working in adiabatic approximation, one supposes
 that $H$ is almost constant and the time-derivatives of $H$ can be
neglected. Then since $a\propto \e^{Ht}$,  using (\ref{CB3}), we find
\be
\label{CB3b}
\rho_A\sim - 24b' H^4\ .
\ee
The first quantum corrected FRW equation looks like
\be
\label{H1}
\frac{6}{\kappa^2}H^2= \rho + \rho_A = \rho - 24 b' H^4\ \ ,
\ee
The above equation can be rewritten as
\be
\label{H2}
0=-24b' \left(H^2 + \frac{1}{4b' \kappa^2}\right)^2 + \frac{3}{2b'\kappa^2} + \rho\ .
\ee
Since $b'$ and therefore $\frac{1}{4b' \kappa^2}$ are negative, in order that $H^2$ has positive real solution,
it follows the constraint for $\rho$
\be
\label{H3}
\rho < - \frac{3}{2b'\kappa^2}\ .
\ee
Thus, even if $\rho$ includes the dark contribution from phantom, $\rho$
 has an upper bound. In other words, it does not grow infinitely with the
time, which was the disaster for phantom cosmology. Equivalent upper bound
may be suggested when one uses Hawking radiation from cosmological horizon
(as it was communicated to us by P. Wang).
Of course, near the Big Rip singularity,
 the time-derivatives of $H$ should be taken into account in the
consistent treatment of the sort presented in the previous section.

Now we consider the Big Rip singularity \cite{CKW} generated by the scalar
field with the exponential potential:
\be
\label{BR1}
S={1 \over \kappa^2}\int d^4 x \sqrt{-g} \left( R - {\gamma \over 2}
\partial_\mu \phi \partial^\mu \phi - V(\phi) \right)\ .
\ee
When $\gamma<0$, the scalar is a phantom with $w<-1$.
By solving the $\phi$-equation of motion
\be
\label{BR2}
0=-\gamma \left(\frac{d^2 \phi}{dt^2} + 3H \frac{d\phi}{dt}\right) - V'(\phi)\ .
\ee
and the first FRW equation
\be
\label{BR3}
{6 \over \kappa^2}H^2=\rho_\phi={\gamma \over 2}\left(\frac{d\phi}{dt}\right)^2 + V(\phi)\ ,
\ee
when
\be
\label{BR4}
V(\phi)=V_0 \e^{-{2\phi \over \phi_0}}\ ,
\ee
one gets a singular solution:
\be
\label{BR5}
\phi=\phi_0\ln \left|\frac{t_s - t}{t_1}\right|\ ,\quad
H=-\frac{\gamma \kappa^2}{4\left(t_s - t\right)}\ ,\quad
t_1^2 \equiv -\frac{\gamma \phi_0^2 \left(1 - \frac{3\gamma \kappa^2}{4}\right)}{2V_0}\ ,
\ee
which gives
\be
\label{BR6}
a=a_0 \left|\frac{t_s - t}{t_1}\right|^{\frac{\gamma \kappa^2}{4}}\ .
\ee
Here $a$ is singular at $t=t_s$ if $\gamma<0$.
General solution of above phantom system has been found in \cite{emilio}.
Even for the general solution, the behavior near $t=t_s$ is not
qualitatively changed from that in (\ref{BR6}). Hence, from the first look
the Big Rip singularity seems to be inevitable.

Near the Big Rip singularity, since $a$ blows up, curvature becomes large
as $R\propto \left|t - t_s\right|^{-2}$.
Since the quantum correction contains powers and higher derivatives of the curvatures in general,
the quantum correction becomes dominant. Hence, one can apply the same
reasoning as in the previous section.
With the account of the quantum correction  (\ref{CB3}),
the corrected FRW equation has
the following form:
\be
\label{BR7}
{6 \over \kappa^2}H^2 ={\gamma \over 2}\left({d\phi \over dt}\right)^2 + V(\phi) + \rho_A\ .
\ee
Let us assume
\be
\label{BR8}
H=h_0 + \delta h \ ,\quad \phi=\phi_0\ln \left|\frac{t_s - t}{t_1}\right| + \delta\phi\ .
\ee
and when $t\to t_s$, $\delta h$, $\delta\phi$ are much smaller than the first terms but
$\frac{d\delta h}{dt}$ can be singular.
Then $\phi$-equation of motion (\ref{BR2}) reduces to
\begin{eqnarray}
\label{BR9}
0&=&-\gamma\left( - \frac{\phi_0}{\left(t_s - t\right)^2} - \frac{3h_0}{t_s - t}\right)
+ \frac{2V_0t_1^2}{\phi_0\left(t_s - t\right)^2}\left(1 - \frac{2}{\phi_0}\delta\phi\right) \nn
&& + o\left(\left(t_s - t\right)^{-1}\right)\ ,
\end{eqnarray}
which gives
\be
\label{BR10}
V_0t_1^2= - {\gamma\phi_0^2 \over 2}\ ,\quad \delta\phi=-\frac{3}{2}\left(t_s - t\right)\ .
\ee
With $\frac{2}{3}b + b''\neq 0$, one gets
\be
\label{BR11}
\rho_A \sim 6h_0 \left(\frac{2}{3}b + b''\right)
\frac{d^2 \delta h}{dt^2}\ .
\ee
Substituting (\ref{BR10}) and (\ref{BR11}) into the quantum corrected FRW
equation
(\ref{BR7}),
we find
\be
\label{BR12}
0=\frac{3\gamma h_0 \phi_0}{t_s - t} + 6h_0 \left(\frac{2}{3}b + b''\right)
\frac{d^2 \delta h}{dt^2} + o\left(\left(t_s - t\right)^{-1}\right)\ ,
\ee
and
\be
\label{BR13}
\delta h=\frac{\gamma\phi_0}{2 \left(\frac{2}{3}b + b''\right)}\left(t_s - t\right)
\ln \left|\frac{t_s - t}{t_2}\right| \ .
\ee
Here $t_2$ is a constant of the integration.
The scale factor $a$ behaves as
\be
\label{BR14}
a=a_0\left|\frac{t_s - t}{t_2}\right|^{\frac{\gamma\phi_0}{4 \left(\frac{2}{3}b + b''\right)}
\left(t_s - t\right)^2}\e^{-h_0\left(t_s - t\right)
 - \frac{\gamma\phi_0}{8 \left(\frac{2}{3}b + b''\right)}\left(t_s - t\right)^2
+ o\left(\left(t_s - t\right)^2\right)}\ .
\ee
There appear logarithmic singularities in $\frac{d^2 a}{dt^2}$, $\frac{dH}{dt}$ but the singularity
 is moderated. Moreover, the universe might develop beyond $t=t_s$.
Thus, quantum effects prevent from the most singular universe.
In case $\frac{2}{3}b + b'=0$, the assumption (\ref{BR8}) seems to be inconsistent.

Another interesting dark energy model (which describes current
acceleration and even current dominance of dark energy) was proposed in
\cite{LR}, where
the matter Lagrangian density (dark energy) is
coupled with the scalar curvature:
\be
\label{LR1}
S=\int d^4 x \sqrt{-g}\left\{{1 \over \kappa^2}R + R^\alpha L_d \right\}\ .
\ee
Here $L_d$ is matter-like Lagrangian density.
The second term  may be induced by quantum
effects as some non-local effective action.
By the variation over $g_{\mu\nu}$, the equation of motion follows:
\be
\label{LR2}
0= {1 \over \sqrt{-g}}{\delta S \over \delta g_{\mu\nu}}
= {1 \over \kappa^2}\left\{{1 \over 2}g^{\mu\nu}R - R^{\mu\nu}\right\}
+ \tilde T^{\mu\nu}\ .
\ee
Here the effective energy momentum tensor (EMT) $\tilde T_{\mu\nu}$ is defined by
\bea
\label{w5}
\tilde T^{\mu\nu} &=& - \alpha R^{\alpha - 1} R^{\mu\nu} L_d  + \alpha\left(\nabla^\mu \nabla^\nu
 - g^{\mu\nu}\nabla^2 \right)\left(R^{\alpha -1 } L_d\right) + R^\alpha T^{\mu\nu}\,\nn
T^{\mu\nu}&\equiv & {1 \over \sqrt{-g}}{\delta \over \delta g_{\mu\nu}}
\left(\int d^4x\sqrt{-g} L_d\right).
\eea
Let free massless scalar be a matter
\be
\label{LR4}
L_d = - {1 \over 2}\partial_\mu \phi \partial_\nu \phi\ .
\ee
The metric (\ref{dSP2}) is chosen.
Assuming $\phi$ only depends on $t$ $\left(\phi=\phi(t)\right)$, the
solution of scalar field equation is given by
\be
\label{LR9}
\dot \phi = q a^{-3} R^{-\alpha}\ .
\ee
Here $q$ is a constant of the integration. Hence $R^\alpha L_d = {q^2 \over 2 a^6 R^\alpha}$,
which becomes dominant when $R$ is small (large) compared with the
Einstein term ${1 \over \kappa^2}R$ if $\alpha>-1$ $\left(\alpha <-1\right)$.


The accelerating solution of FRW equation exists \cite{LR}
\bea
\label{LR13}
&& a=a_0 t^{\alpha + 1 \over 3}\quad \left(H={\alpha + 1 \over 3t}\right)\ ,\nn
&& a_0^6 \equiv {\kappa^2 q^2 \left(2\alpha - 1\right)\left(\alpha - 1\right)
\over 3\left(\alpha + 1\right)^{\alpha + 1}
\left({2 \over 3}\left(2\alpha - 1\right)\right)^{\alpha + 2}}\ .
\eea
Eq.(\ref{LR13}) tells that the universe accelerates, that is, $\ddot a>0$ if $\alpha>2$.

For the matter with the relation $p=w\rho$, where $p$ is the pressure and $\rho$ is the
energy density, from the usual FRW equation, one has $a\propto t^{2 \over 3(w+1)}$.
For $a\propto t^{h_0}$ it follows $w=-1 + {2 \over 3h_0}$,
and the accelerating expansion ($h_0>1$) of the universe occurs if $w<-{1 \over 3}$.
For  (\ref{LR13}), one gets
\be
\label{LR13b}
w={1 - \alpha \over 1 + \alpha}\ .
\ee
Then if $\alpha<-1$,  $w<-1$, i.e. an effective phantom.

When $\alpha<-1$, i.e., $w<-1$, the universe is shrinking in the solution
(\ref{LR13}). However, if one changes the
direction of time as $t\to t_s - t$, the universe is expanding but has a
Big Rip singularity at $t=t_s$. Since near the singularity, the curvature becomes very large again,
we may include the quantum correction (\ref{CB3}) 
\be
\label{LRLR1}
0=-{3 \over \kappa^2}H^2 + \tilde \rho + \rho_A \ .
\ee
If $H$ behaves as in (\ref{LR13}), after changing the direction of the time as $t\to t_s - t$ in $a$,
\be
\label{LRLR2}
H=\frac{1}{a}\frac{da}{dt}=\frac{\alpha + 1}{3\left(t-t_s\right)}\ ,
\ee
the quantum correction of the energy density $\rho_A$ behaves as $\rho_A\sim \left(t-t_s\right)^{-4}$,
which becomes very large when $t\sim t_s$. This shows that $H$ cannot
grow as in (\ref{LRLR2}).
If $H$ is not very large, $\tilde \rho$  (\ref{LRLR1})
becomes very small
when $a$ is large and can be neglected since $\tilde \rho \propto a^{-6}$.
 In such a situation,
Eq.(\ref{LRLR2}) reduces to (\ref{CB6}). Hence, instead of future
singularity, due to quantum effects the dark energy universe ends up in
the deSitter phase
(\ref{CB10}). Thus, quantum effects resolve the sudden future singularity
of dark energy universe.

\section{Thermodynamics and entropy bounds in the dark energy universe
\label{entropy}}

Thermodynamics of dark energy universe was discussed in ref.\cite{BNOV}
where the appearence of negative entropies for models with equation of
state parameter less than $-1$ was demonstrated and entropy bounds were
constructed. In the present section  the entropy bounds near
the Big Rip singularity are considered.
The Hubble entropy $S_H$, Bekenstein entropy $S_B$, and Bekenstein-Hawking entropy $S_{BH}$ are
defined by
\be
\label{Eb1}
S_H=\frac{HV}{2G}\ ,\quad S_B = \frac{2\pi a E}{3}\ ,\quad S_{BH}=\frac{V}{2Ga}\ .
\ee
Here $G$ is a gravitational constant ($\kappa^2=16\pi G$) and $V$ is
the volume of the universe
where for the universe with flat spatial part, it is chosen
\be
\label{Eb2}
V=V_0a^3\ .
\ee
Near the Big Rip singularity, the dark energy dominates
and the usual matter contribution may be neglected.
Without quantum correction,  $a\sim \left(t_s - t\right)^{\frac{2}{3(w+1)}}$ and
$\rho\sim a^{-3\left(1+w\right)} \sim \left(t_s - t\right)^{-2}$
in accord with (\ref{BR5}) and (\ref{BR6}).
The entropies behave as
\be
\label{Eb3}
S_B \sim \left(t_s - t\right)^{\frac{2(1-3w)}{3(w+1)}}\ , \quad
S_H \sim \left(t_s - t\right)^{\frac{1-w}{w+1}}\ ,\quad
S_{BH} \sim \left(t_s - t\right)^{\frac{4}{3(w+1)}} \ ,
\ee
where the exponents are related by
\be
\label{Eb4}
\frac{2(1-3w)}{3(w+1)}<\frac{1-w}{w+1}<\frac{4}{3(w+1)}<0\ ,
\ee
when $w<-1$. Hence, all the entropies are singular at $t=t_s$.
Eq.(\ref{Eb4}) shows that
the Bekenstein entropy $S_B$ is most singular
 while the Bekenstein-Hawking entropy is less singular.

In order to estimate the entropy, we consider the thermodynamical model \cite{BNOV}, where
the free energy corresponding to matter with $w$ is given by
\be
\label{Eb5}
F_w=T\hat F\left(T^{\frac{1}{w}}V\right)\ .
\ee
Here $T$ is the temperature and $V$ is the volume of the system.
$\hat F$ is a function determined by the matter.
The thermodynamical parameters are
\bea
\label{Eb6}
p&=& - \frac{\partial F_w}{\partial V}= - T^{1+ \frac{1}{w}}{\hat F}'\left(T^{\frac{1}{w}}V\right)\ , \nn
\rho &=& \frac{1}{V}\left(F_w - T\frac{\partial F_w}{\partial T}\right)
= - \frac{1}{w}T^{1+ \frac{1}{w}}{\hat F}'\left(T^{\frac{1}{w}}V\right)\ , \nn
{\cal S}&=& - \frac{\partial F_w}{\partial T}
= - {\hat F}\left(T^{\frac{1}{w}}V\right) - \frac{1}{w}T^{\frac{1}{w}}{\hat F}'\left(T^{\frac{1}{w}}V\right)\ .
\eea
Here ${\cal S}$ is an entropy.
Since $\rho$ behaves as $\rho=\rho_0 a^{-3\left(1+w\right)} \propto \rho_0\left(\frac{V}{V_0}\right)^{-(1+w)}$,
 from the second equation (\ref{Eb6}) it follows
\be
\label{Eb7}
\rho_0 {V_0}^{1+w}\propto  - \frac{1}{w}\left(T^{\frac{1}{w}}V\right)^{1+w}{\hat F}'\left(T^{\frac{1}{w}}V\right)\ .
\ee
Then  $T^{\frac{1}{w}}V$ should be a constant, which indicates that the
entropy ${\cal S}$ in the third
equation  (\ref{Eb6}) is also a constant. Since  Hubble, Bekenstein, and Bekenstein-Hawking
entropies  (\ref{Eb3}) diverge at the Big Rip singularity,
the following entropy bound holds near the (classical)
Big Rip singularity
\be
\label{Eb8}
{\cal S}<S_{BH}<S_H<S_B\ .
\ee
Here Eq.(\ref{Eb4}) was used. Then all the bounds are satisfied, which may
be compared with the case of the
brane-world dark energy model of ref.\cite{abdalla}, where the Hubble
entropy bound is not satisfied and
the Bekenstein bound is often violated.

The Hubble parameter in the expanding universe is given by
$H={\frac{2}{3(w+1)}}{t}$ if $w>-1$ and $H={\frac{2}{3(w+1)}}{t_s - t}$ if $w<-1$,
$\frac{dH}{dt}<0$ when $w>-1$ and $\frac{dH}{dt}>0$ if $w<-1$, which corresponds to the
Big Rip singularity.
  $S_B\propto H^2 V a \propto H^2 a^4$ as found from the FRW equation.
Then one may define the
following quantity (which may indicate the future singularity
occurence)
\be
\label{Ebb1}
\tilde S\equiv \frac{S_{BH}^2}{S_B}\ ,
\ee
where  $\tilde S \propto H^{-2}$ since $S_{BH}\propto a^2$. Hence, if
$\tilde S$ decreases with time,
there might occur the Big Rip singularity.

Let us reconsider the above entropy bounds with the account of the quantum
effects. Using the solution
(\ref{BR8}), (\ref{BR10}), (\ref{BR13}), (\ref{BR14}),
 we find the entropies  (\ref{Eb3}) behave as
\be
\label{Eb9}
S_H\to  \frac{h_0V_0a_0^3}{2G}\ ,\quad S_B\to \frac{2\pi}{3}\frac{2\gamma h_0 \phi_0 V_0a_0^4}{t_s - t} \ ,
\quad S_{BH}\to \frac{V_0 a_0^2}{2G}\ .
\ee
Then $S_H$ and $S_{BH}$ are finite and they may give meaningful entropy
bound but
$S_B$ is negative since $\gamma<0$ and diverges. Hence, the Bekenstein
bound $S<S_B$
is violated. In (\ref{Eb9}), however, we have  included only the classical
part $\rho_\phi$
(\ref{BR3}) in order to estimate $S_B$. With the account of the quantum
correction $\rho_A$  (\ref{CB3}),
the singularity in $S_B$ can be cancelled. Since $\rho=\rho_\phi + \rho_A=\frac{6}{\kappa^2}H^2
=\frac{3H^2}{8\pi G}$, we find the following expression of the quantum corrected Bekenstein entropy $S_B^q$:
\be
\label{Eb10}
S_B^q =  \frac{h_0^2V_0a_0^4}{4G}\ ,
\ee
which is positive and finite. With $S_B^q$ (\ref{Eb10}) instead of
$S_B$  (\ref{Eb9}), all the
entropies are finite. We should note
\bea
\label{Eb11}
& S_B^q\gg S_H\gg S_{BH} \quad & \mbox{if} \quad h_0\gg \frac{1}{a_0}\ ,\nn
& S_B^q\ll S_H\ll S_{BH} \quad & \mbox{if} \quad h_0\ll \frac{1}{a_0}\ .
\eea
The parameters $h_0$ and $a_0$ are the values of the Hubble parameter and the size of universe at $t=t_s$,
which may be determined from the proper initial conditions.

In \cite{ebNO}, the (quantum corrected) entropy bounds  have been
discussed. In \cite{ebNO},
the spatial part of the universe is sphere, where we have a relation
\cite{verlinde}
\be
\label{Eb12}
S_H^2 + \left(S_{BH} - S_B\right)^2 = S_B^2\ ,
\ee
even with the quantum correction. In case that the spatial part is flat,
Eq.(\ref{Eb12}) reduces to
$S_H^2=2S_{BH}S_B$. We should note that for $S_H$, $S_{BH}$ (\ref{Eb9})
and $S_B^q$  (\ref{Eb10}),
it holds
\be
\label{Eb13}
S_H^2=2S_{BH}S_B^q\ .
\ee
Even for the classical case that all the entropies are singular,
 $S_H^2=2S_{BH}S_B$, which can be
found from the FRW equation (\ref{BR3}). One can rewrite the FRW equation
(\ref{BR3}) in the form
$S_H^2=2S_{BH}S_B$ by using the definition of the entropies (\ref{Eb3}).

To conclude, it is shown that entropies near to classical singularity are
singular as well. However, quantum corrected entropies are  finite
and give the well-defined entropy bounds. This is not surprised due to
the fact that quantum effects help to escape of future singularity
in the dark energy universe.

\section{Dark energy universe with general equation of state
\label{TW}}

So far we concentrated mainly on the various aspects of the dark energy
universe with negative equation of state parameter which is less than
$-1$. Nevertheless, the recent astrophysical data admit also the case
of time-dependent equation of state parameter. Let us consider
several examples of such dark energy cosmology and its late time
behaviour. Note that several models of dark energy universe with
time-dependent  equation of state were discussed in
\cite{BNOV,time} (see also refs. therein).

One starts from the general equation of state of the form
\be
\label{TW1}
p=f(\rho)\ ,
\ee
instead of the equation $p=w\rho$ with constant $w$. In (\ref{TW1}), $f$ can be an arbitrary function.
Imagine that solving gravitational equations, we want to construct the
cosmology
with time-dependent $w$, which describes the transition from the
deccelerating universe
 to the accelerating one. As an example,  the following scale factor
$a(t)$ may be considered
\be
\label{TW2}
a=a_0\e^{\lambda t} t^\alpha\ .
\ee
Here $\lambda$ and $\alpha$ are some constants. Hence,
\bea
\label{TW3}
\frac{da}{dt}=a H &=& a\left(\lambda + \frac{\alpha}{t}\right)\ ,\nn
\frac{d^2 a}{dt^2} &=& a\left\{\left(\lambda + \frac{\alpha}{t}\right)^2 - \frac{\alpha}{t^2}\right\}\ .
\eea
In the case that $\lambda$ and $\alpha$ are positive, the universe is
accelerating if
\be
\label{TW4}
t>t_0\equiv \frac{\sqrt{\alpha} - \alpha}{\lambda}\ ,
\ee
and deccelerating if
\be
\label{TW5}
t<t_0\ .
\ee
That is, the deccelerating universe turns into the accelerating one at $t=t_0$.
 Thus, if transition point $t_0$ occured
about 5 billion years ago, the solution may approximately describe  our
universe.
Note that $t_0$ is positive when
$0<\alpha<1$.

By using (\ref{F1}), one finds
\bea
\label{TW6}
\rho&=&\frac{6}{\kappa^2}\left(\lambda + \frac{\alpha}{t}\right)^2\ ,\\
\label{TW7}
p&=&-\frac{2}{\kappa^2}\left(\frac{\alpha(3\alpha - 2)}{t^2} + \frac{6\lambda\alpha}{t} + 3\lambda^2\right)\ .
\eea
Eq.(\ref{TW6}) can be solved as
\be
\label{TW8}
t=\frac{\alpha}{\frac{\kappa^2\rho}{6} - \lambda}\ ,
\ee
substituting (\ref{TW8}) into (\ref{TW7}), we obtain
\be
\label{TW9}
p=-\frac{2}{\kappa^2}\left\{ \left(3 - \frac{2}{\alpha}\right)\left(\frac{\kappa^2\rho}{6}\right)^2
+ \frac{4}{\alpha}\frac{\kappa^2\rho}{6} - \left(6 - \frac{2}{\alpha}\right)\lambda^2\right\}\ .
\ee
Hence, with (\ref{TW9}) as the equation of state, we arrive at a solution
 (\ref{TW2}), where the deccelerating universe turns into the accelerating
 one.

General case  (\ref{TW1}) may be considered as well.
Using the first FRW equation (\ref{B2}) and the energy conservation law
\be
\label{TW11}
0=\frac{d\rho}{dt} + 3 H\left(\rho + p\right)\ ,
\ee
one gets
\be
\label{TW15}
\frac{d\rho}{dt}=F(\rho)\equiv - \kappa\sqrt{\frac{3\rho}{2}}\left(\rho + f(\rho)\right)\ .
\ee
With a proper assumption about function $f(\rho)$, we can find the
$t$-dependence of $\rho$ by solving Eq.(\ref{TW15}).
Using the obtained expression for $\rho=\rho(t)$, one can also find the
$t$-dependence of $p$ as
$p=f\left(\rho(t)\right)$.

By combining (\ref{TW1}) and (\ref{TW15}), the pressure $p$ can be expressed as
\be
\label{TW16}
p=-\rho - \frac{1}{\kappa}\sqrt{\frac{2}{3\rho}}F(\rho)\ .
\ee
Therefore if $F(\rho)>0$ $\left(F(\rho)<0\right)$, it follows $w<-1$
$\left(w>-1\right)$.
We now assume $F(\rho)=0$ at $\rho=\rho_0$, where $\rho_0$ is a particular
value of $\rho$.
We further assume that when $\rho\sim \rho_0$, $F(\rho)$ behaves as
\be
\label{TW17}
F(\rho)\sim F_0\left(\rho - \rho_0\right)^n\ .
\ee
Here $F_0$ is a constant and $n$ is a positive odd integer.
If $n\neq 1$, by solving (\ref{TW15}), one gets
\be
\label{TW18}
\rho \sim \rho_0 + \left\{F_0\left(1-n\right)\left(t-t_0\right)\right\}^{-\frac{1}{n-1}}\ ,
\ee
and if $n=1$,
\be
\label{TW19}
\rho \sim \rho_0 + C\e^{F_0 t}\ .
\ee
Here $t_0$ or $C$ is a constant of the integration. Then $\rho$ goes to $\rho_0$ only at $|t|\to \infty$,
which may indicate that the region with $w>-1$ could be disconnected with
the region $w<-1$.
Instead of  a positive odd integer $n$, one may start from
\be
\label{TW20}
n=\frac{m-1}{m}\ ,
\ee
with an integer $m$. Then as in (\ref{TW18}) the time-dependent energy
density looks like
\be
\label{TW21}
\rho \sim \rho_0 + \left\{\frac{F_0}{m}\left(t-t_0\right)\right\}^m\ .
\ee
Thus, the region $w>-1$ might be connected with the region $w<-1$. In
this case, however, the equation of state has branches.


We now consider the case with a linear equation of state $p=w\rho$ where
$w$ depends on time as $w=w(t)$. Replacing $f(\rho)$ by $w(t)\rho$ in
(\ref{TW15}) and using 
 the first FRW equation (\ref{B2}), it follows
\be
\label{TW15c}
H=\frac{2}{3}\left(\int \left(1+w(t)\right)\right)^{-1}\ .
\ee
In order to investigate what  happens when $w$ changes the value from the
one bigger than $-1$ to that less than
$-1$, we now assume that near $t=t_0$ $w(t)$ behaves as
\be
\label{TW16c}
w(t)\sim -1 + w_0\left(t-t_0\right)\ ,
\ee
with constant $w_0$. Using
\be
\label{TW16b}
\int dt \left( 1+w(t) \right) \sim \frac{1}{2}w_0\left(t-t_0\right)^2 + w_1\ ,
\ee
one finds
\be
\label{TW17c}
\rho\sim \frac{8}{3\kappa^2\left\{w_0\left(t-t_0\right)^2 + 2w_1\right\} }\ ,\quad
H\sim \frac{32}{3\left\{w_0\left(t-t_0\right)^2 + 2w_1\right\} }\ .
\ee
Here $w_1$ is a constant of the integration. Unlike  the case in
(\ref{TW18}), there is no singularity
at $t=t_0$ if $w_1\neq 0$.

As one more example,  the case with another $w(t)$ may be considered
\be
\label{TW18c}
w(t)=-1 -\frac{a\left(t-t_0\right)}{t + b}\ .
\ee
Here $a$, $b$, and $t_0$ are positive constants.
Then $w(t)$ has the following properties
\bea
\label{TW19c}
&& w(0)=w_0\equiv -1 + \frac{at_0}{b} > -1\ ,\quad w\left(t_0\right)=-1\ ,\nn
&& w(+\infty)=w_\infty \equiv -1 - a < -1\ .
\eea
Hence, $w(t)$ connects the region of $w>-1$ with that of $w<-1$.
Since
\be
\label{TW20c}
W(t)\equiv \int dt \left( 1+w(t) \right) = a\left\{ -t + \left( b+t_0 \right)\ln \frac{t + b}{t_1}\right\}\ .
\ee
we find
\be
\label{TW21c}
\rho(t)=\frac{8}{3\kappa^2 W(t)^2}\ ,\quad H=\frac{2}{3W(t)}\ .
\ee
where $t_1$ is a constant of the integration.
When $t\sim t_0$, one gets
\be
\label{TW22c}
W(t)\sim a\left( - t_0 + \left(b+t_0\right) \ln \frac{t_0 + b}{t_1}\right)
+ \frac{a\left(t-t_0\right)^2}{2\left(b+t_0\right)} + {\cal O}\left(\left(t-t_0\right)^2\right)\ ,
\ee
which is consistent with (\ref{TW16b}).
If the universe is expanding, that is $H>0$, at $t=0$ we find the following condition
\be
\label{TW24c}
b>t_1>0\ .
\ee
$W(t)$ behaves as $W(t)\sim - at<0$ when $t$ is large. Thus, if the
condition (\ref{TW24c}) is satisfied,
$W(t)$ vanishes at finite $t$ ( $t=t_s$) where $t_s$ is a solution of the
equation
\be
\label{TW25c}
0=W\left(t_s\right) =a\left\{ -t_s + \left( b+t_0 \right)\ln \frac{t_s + b}{t_1}\right\}\ .
\ee
Hence, there appears  singularity at $t=t_s>0$ in $\rho$ and $H$, which is
nothing but the Big Rip singularity.
However, even with (\ref{TW24c}), since
$t_s - t_0 = \left( b+t_0 \right)\ln \frac{t_s + b}{t_1} - t_0$ can be negative in general, 
the singularity may occur in the region $w>-1$.

To conclude, we presented several examples of dark energy universe with
time-dependent (negative) equation of state. The possibility to have
naturally accelerated universe phase (sometimes, as a transition from
decceleration) is shown. It is interesting that when time-dependent
equation of state parameter is negative (not only less but even bigger
than $-1$) the finite time future singularity occurs as a final state of
such universe. Nevertheless, in the same way as it was discussed in
second and third sections one can show that quantum effects prevent
the evolution to such final state( eventually driving the universe
to the inflationary era).

\section{Discussion}

In summary, we discussed several aspects of phantom thermodynamics
and the final state of the phantom dark energy universe. Despite
the absence of consistent phantom energy theory, some general results
look quite promising. In particulary, it is shown that finite time Big Rip
singularity remains to be deeply theoretical possibility in classical
phantom theory. The account of quantum effects (when universe evolves to
the singularity and when curvature invariants grow) is done. As a result,
it is proved that quantum effects moderate the singularity  or it
 even disappears
completely. (Note also that stability analysis\cite{emilio}, and gravitational perturbations account\cite{yu} indicates that that perturbations act 
against the Big Rip occurence.) Hence, it is unlikely that the final state 
of phantom universe
is Big Rip. Rather, the final state is the initial state on the same time,
because
 inflationary era may start again in the future. The resolution of Big Rip
singularity resolves also several related phenomena. For instance, entropy
bounds which are divergent near Big Rip become well-defined after the
quantum corrections are included. Similarly, escape of finite time
singularity means that black holes mass evolution is less dramatic than
it was predicted (masses do not vanish to zero).

It is expected that soon precise observational cosmology data will give
more stringent bounds for equation of state parameter. At the moment,
it is still unclear if it will lie at quintessence, or at phantom region.
Moreover, it is quite possible that the smiling universe hides a number of
surprises for us. Nevertheless, phantom universe remains to be the
theoretical possibility which is not explored yet and which deserves some
attention.

\section*{Acknowledgments}

S.D.O. would like to thank M.C. Abdalla for kind hospitality at IFT, Sao
Paulo. This research has been supported in part by the Ministry of
Education,
Science, Sports and Culture of Japan under grant n.13135208 (S.N.),
by  grant 2003/09935-0 of FAPESP, Brazil (S.D.O.) and by project
BFM2003-00620, Spain (S.D.O.).

\appendix

\section{The entropy of phantom universe }

One more weird property of the dark energy universe with $w$ less than
$-1$ is the strange behavior of the entropy. In fact, it was pointed out
in ref.\cite{BNOV} that entropy of such universe is negative.
Another proposal came out in ref.\cite{Pedro} suggesting to consider
phantom fluid as kind of cosmological quantum fliud (as nuclear spin
model, for instance) where negative temperature is admitted. (Note 
that the idea of negative temperature in cosmological context 
was discussed first by Vanzo-Klemm \cite{KV}).
In this case,
the entropy may be positive.

Let us describe the relation between the entropy and energy of such dark
energy universe when the temperature is negative. Starting from the model
\cite{BNOV}, instead of (\ref{Eb5}) we consider the following free
energy:
\be
\label{P1}
F_w=\gamma T\tilde F\left(\left(\gamma T\right)^{\frac{1}{w}}V\right)\ .
\ee
If the temperature $T$ is positive,  $\gamma=1$ and if it is negative,
$\gamma=-1$.
Simple calculation gives the pressure $p$, the energy density $\rho$, and
the
entropy ${\cal S}$:
\bea
\label{P2}
p&=& - \left(\gamma T\right)^{1+ \frac{1}{w}}{\tilde F}'\left(\left(\gamma T\right)^{\frac{1}{w}}V\right)\ , \nn
\rho &=& - \frac{1}{w}\left(\gamma T\right)^{1+ \frac{1}{w}}{\tilde F}'\left(\left(\gamma T\right)^{\frac{1}{w}}V\right)\ , \nn
{\cal S}&=& -\gamma\left\{ {\tilde F}\left(\left(\gamma T\right)^{\frac{1}{w}}V\right)
+ \frac{1}{w}T^{\frac{1}{w}}{\tilde F}'\left(\left(\gamma T\right)^{\frac{1}{w}}V\right)\right\}\ .
\eea
If the energy is extensive,  the energy behaves as $E=\rho V\to \lambda E$ under the rescaling the entropy
and the volume as ${\cal S}\to \lambda {\cal S}$ and $V\to \lambda V$.
In accord with \cite{BNOV} we consider the following free energy:
\be
\label{P3}
F_w=-f_0 \left(\gamma T\right)^{1+{1 \over w}}V\left(1 + f_1 \left(\gamma T\right)^{-{2 \over nw}}
V^{-{2 \over n}}\right)\ .
\ee
If there is no the second term, the first term gives the extensive energy.
It is assumed the second term is small compared with the first term.
Then one gets
\bea
\label{P4}
E&=&\frac{pV}{w} \nn
&=&{f_0 \over w}\left(\gamma T\right)^{1+{1 \over w}}V\left(1 +
\left(1-{2 \over n}\right)f_1 \left(\gamma T\right)^{-{2 \over nw}}V^{-{2 \over n}}\right)\ ,\nn
{\cal S}&=&f_0 \gamma \left(\gamma T\right)^{1 \over w}V\left(\left(1 + {1 \over w}\right) \right.\nn
&& \left. + \left(1 + {1 \over w} - {2 \over nw}\right)f_1 \left(\gamma T\right)^{-{2 \over nw}}V^{-{2 \over n}}\right)\ .
\eea
The sub-extensive part of the energy $E_C$, which is called the Casimir
energy \cite{verlinde}, is given by
\bea
\label{P5}
E_C &=& n\left(E+pV-TS\right)=-nV^2{\partial \over \partial V}\left({F \over V}\right) \nn
&=& -2 f_0f_1 \left(\gamma T\right)^{1+{1 \over w} - {2 \over nw}}V^{1 - {2 \over n}}\ .
\eea
The extensive part of the energy $E_E$ has the following form:
\bea
\label{P6}
E_E&=&E-{1 \over 2}E_C \nn
&=& {f_0 \over w} \left(\gamma T\right)^{1+{1 \over w}}V\left(1 +
\left(1-{2 \over n} + w\right)f_1 \left(\gamma T\right)^{-{2 \over nw}}V^{-{2 \over n}}\right)\ .
\eea
As in Section \ref{entropy}, $T^{\frac{1}{w}}V$ is a constant in the phantom dominated universe.
Then if one neglects the second term in $E_E$ and/or $E$ as
\be
\label{P7}
E_E\sim E \sim {f_0 \over w} \left(\gamma T\right)^{1+{1 \over w}}\ ,
\ee
we obtain
\be
\label{P8}
S\sim A\left[V^w \sqrt{\left(2E-E_C\right)E_C}\right]^{n \over (w+1)n-1} \ .
\ee
Here $A$ is a constant.

The natural assumption is $E>0$. From the expression  (\ref{P4}),
 $f_0<0$ if $w<0$.
 Let the starting condition is that the entropy ${\cal S}$ is positive.
In case of the quintessence, where $-1<w<\frac{1}{3}$, since $1+\frac{1}{w}<0$,
 from Eq.(\ref{P4}), we find $\gamma>0$ (positive temperature) so that
the entropy is positive.
On the other side, in case of  phantom, where $w<-1$, that is,
$1+\frac{1}{w}>0$, it follows $\gamma<0$
if the entropy ${\cal S}$ is positive. Therefore the temperature should be
 negative. Conversely,
if we assume the temperature is positive in the phantom theory,
 the entropy should be negative.

Note also that in order to obtain Cardy-Verlinde formula (\ref{P8})
(for list of references, see review \cite{NOOfrw}),  the
Casimir energy $E_C$ should be positive, which
requires $f_1>0$. Hence, the entropy of phantom-filled universe is
positive when the temperature is negative. In this case, standard CV
entropy formula holds.

\section{Black holes mass evolution in the dark energy universe}

One more strange feature of the phantom universe is the black holes mass
loss up to the full disappearence in the Big Rip singularity. The
corresponding analysis \cite{BDE} was performed in classical phantom-like
universe (where dominant energy condition is broken) with final state in
the Big Rip. In the present Appendix,
we reconsider this process taking into account the quantum effects
which prevent the creation of Big Rip singularity as well as
time-dependent (negative) equation of state.

As is shown in an important paper \cite{BDE} (see also \cite{Pedro}), the rate of the black
hole mass change in the fluid
with the energy density $\rho$ and the pressure $p$ is given by
\be
\label{AC1}
\frac{dM}{dt}=4\pi A M^2 \left(\rho + p\right)\ .
\ee
Here $M$ is the mass of the black hole and $A$ is a dimensionless positive constant.
As a background,  FRW universe with the metric (\ref{dSP2}) may be
considered.
Combining the first FRW equation (\ref{B1}) and the energy conservation
law (\ref{TW11}), one obtains
\be
\label{AC3}
\rho + p = -\frac{2}{\kappa}\sqrt{\frac{2}{3}}\frac{d\left(\rho^{\frac{1}{2}}\right)}{dt}\ .
\ee
Further combining (\ref{AC1}) and (\ref{AC3}), we get
\be
\label{AC4}
\frac{d}{dt}\left(\frac{1}{M}\right)=\frac{8\pi A}{\kappa}\sqrt{\frac{2}{3}}\frac{d\left(\rho^{\frac{1}{2}}\right)}{dt}\ .
\ee
The solution of above equation is:
\be
\label{AC5}
M=\frac{M_0}{1+ \frac{8\pi AM_0}{\kappa}\sqrt{\frac{2\rho}{3}}}\ .
\ee
Hence, if $\rho$ increases as in the case that the fluid is phantom, $M$
decreases. At the Big Rip
singularity where $\rho$ diverges, $M$ vanishes. This is an universal
property
for any black hole in such phantom universe.
On the other hand, in the case of  Barrow model where $\rho$
is finite (\ref{B3}), the mass $M$ is finite even at the singularity.
By using the first FRW equation  (\ref{B1}), we may further rewrite
  $M$ (\ref{AC5})
in the following form
\be
\label{AC5b}
M=\frac{M_0}{1+ \frac{4\pi AM_0}{3} H}\ .
\ee

In \cite{BDE,Pedro}, the behavior of $M$ for the phantom with constant $w<-1$
has been investigated
in detail. As in (\ref{BR5}), when $w<-1$ in the expanding universe, the Hubble parameter behaves as
\be
\label{BR5b}
H=-\frac{\gamma \kappa^2}{4\left(t_s - t\right)}=\frac{-\frac{2}{3(w+1)}}{t_s - t}\ .
\ee
On the other hand, in case $w>-1$ it looks like
\be
\label{AC5c}
H=\frac{\frac{2}{3(w+1)}}{t}\ .
\ee
Using Eq.(\ref{AC5b}) one arrives at
\be
\label{AC5d}
M=\frac{M_0}{1 - \frac{4\pi AM_0}{3} \frac{\frac{2}{3(w+1)}}{t_s - t}}\ .
\ee
when $w<-1$ and
\be
\label{AC5e}
M=\frac{M_0}{1 + \frac{4\pi AM_0}{3} \frac{\frac{2}{3(w+1)}}{t}}\ .
\ee
when $w>-1$. In case of (\ref{AC5d}), $M$ decreases and vanishes at $t=t_s$. Near $t=t_s$, $M$ behaves as
\be
\label{AC5f}
M\sim - \frac{9(w+1)\left(t_s - t\right)}{8\pi A}\ .
\ee
This does not depend on $M_0$ and is universal as pointed out in
\cite{BDE}.
On the other hand, $M$ (\ref{AC5e}) increases and reaches the maximal value $M=M_0$ when $t=\infty$.
Even with the account of  cosmological term, the
qualitative behavior does not change.

The above behavior is modified when quantum effects are taken into account
because as it was argued in second and third sections they may stop
the evolution to final singularity. Indeed, let us
 consider the case that the quantum correction is included as in (\ref{BR7}).
If $\frac{2}{3}b + b''\neq 0$,  combining (\ref{BR8}) and (\ref{BR13}),
 it follows that $H=h_0$ when $t=t_s$.
Therefore $M$ has a finite, non-vanishing value:
\be
\label{AC5g}
M\to \frac{M_0}{1+ \frac{4\pi AM_0}{3} h_0}\ .
\ee
Since
\be
\label{AC5h}
\frac{dH}{dt}=\frac{d\delta h}{dt}
=-\frac{\gamma\phi_0}{2 \left(\frac{2}{3}b + b''\right)}\left(
\ln \left|\frac{t_s - t}{t_2}\right| +1\right)\ .
\ee
and $\gamma<0$ for the phantom, near the singularity $t\sim t_s$, $\frac{dH}{dt}>0$
$\left(\frac{dH}{dt}<0\right)$ if $\frac{2}{3}b + b''<0$ $\left(\frac{2}{3}b + b''>0\right)$.
Therefore if $\frac{2}{3}b + b''<0$, since $H$ increases, $M$ decreases
towards the singularity
although $M$ is finite and non-vanishing there. On the other hand, if $\frac{2}{3}b + b''<0$, $M$ increases.
As the Hawking radiation occurs due to the quantum correction,
 the above type of behavior may be more realistic in the phantom universe
(quantum effects have been neglected, at least in the leading-order, in
\cite{BDE}).

Let us reconsider  what happens with the black hole mass
$M$ evolution equation (\ref{AC5}) in the dark energy universe with
time-dependent equation of state.
For simplicity, the quantum corrections are neglected.
When $p=w(t)\rho$, from (\ref{AC3}) it follows
\be
\label{AC6}
\left(1+w(t)\right)\rho  = -\frac{2}{\kappa}\sqrt{\frac{2}{3}}\frac{d\left(\rho^{\frac{1}{2}}\right)}{dt}\ .
\ee
First we consider the case  (\ref{TW16c}) and the behavior of $\rho$ (\ref{TW17c}).
Then if $w_0$ and $w_1$ are positive, $\rho$ takes a minimum value at
$t=t_0$. The black hole mass $M$ (\ref{AC5}) increases when $t<t_0$ and it
reaches the maximum at $t=t_0$.
When $t>t_0$, the mass decreases.

As a more concrete example,  $w(t)$ (\ref{TW18c}) may be discussed.
When $t_s>t_0$ (at $t=t_s$, $W(t)$ (\ref{TW25c}) vanishes), $W(t)$
increases when $t<t_0$ and
decreases when $t>t_0$.
Then the energy density $\rho$ decreases when $t<t_0$, increases when $t>t_0$, and diverges at $t=t_s$.
Therefore the behavior of the black hole mass $M$ (\ref{AC5}) is similar to $W(t)$, that is,
$M$ increases when $t<t_0$, decreases when $t>t_0$ and vanishes at
$t=t_s$ (like in classical phantom universe). Nevertheless, the account of
quantum effects, as we showed above, qualitatively changes the black hole
mass evolution. In other words, the same phenomenon which drives the dark
energy universe out of final singularity (because of second quantum
gravity
era) is responsible for much less sharp loss of black holes
masses.
As Big Rip does not occur, initially massive black holes continue to be
(may be less) massive!

\end{document}